\documentclass[twocolumn,pra,superscriptaddress]{revtex4-2}
\usepackage{graphicx,subfigure}          
\usepackage{amsfonts,amsmath,amssymb,amsthm}    
\usepackage{dsfont}                      
\usepackage[dvipsnames]{xcolor}          
\usepackage[
colorlinks=true,
linkcolor=blue,
citecolor=blue,
urlcolor=blue]{hyperref}                
\usepackage{comment}                    
\usepackage{physics}                    
\usepackage{bm}                         
\usepackage{mdframed}                   
\usepackage{enumitem}                   
\usepackage{soul}                       
\usepackage{svg}

\begin{document}

\title{Efficient circuit compression by multiqudit entangling gates in linear optical quantum computation}
\author{Apurav Tehri}
\email{d23185@students.iitmandi.ac.in}
\affiliation{Center for Quantum Science and Technologies, Indian Institute of Technology, Mandi, Himachal Pradesh 175075, India}
\author{Jaskaran Singh}
\email{jaskaran@iitmandi.ac.in}
\affiliation{Center for Quantum Science and Technologies, Indian Institute of Technology, Mandi, Himachal Pradesh 175075, India}

\begin{abstract}

Linear optical quantum computation (LOQC) offers a promising platform for quantum information processing, but its scalability is fundamentally constrained by the probabilistic nature of non-local entangling gates. Qudit circuit compression schemes mitigate this issue by encoding multiple qubits onto qudits. However, these schemes become inefficient when only a subset of the encoded qubits is required to participate in the non-local entangling gate, leading to an exponential increase in the number of non-local gates. In this paper, we address this bottleneck by demonstrating the existence of multi-level control-Z (CZ) operations for qudits encoded in multiple spatial modes in LOQC. Unlike conventional two-level CZ gates, which act only on a single pair of modes, multi-level CZ gates impart a conditional phase shift for an arbitrarily chosen subset of the spatial modes. We present two explicit linear optical schemes that realize such operations, illustrating a fundamental trade-off between prior information about the input quantum state and the physical resources required. The first scheme is realized with a constant success probability of $1/8$ independent of the qudit dimension using a single non-local entangling operation, at the cost of input state dependence. Our second scheme provides a fully state independent realization, akin to a standard quantum gate, reducing the number of non-local gates to $\mathcal{O}(2^{r_1}+2^{r_2})$ as compared to the existing bound of $\mathcal{O}(2^{r_1+r_2})$ where $r_1$ and $r_2$ are the number of qubits to be removed as control in the qudits. The success probability of the realization is $\frac{1}{2} \left(\frac{1}{8}\right)^{2^{r_1}+2^{r_2}}$. When combined with qudit circuit compression schemes, our results improve upon a key scalability limitation and significantly improve the efficiency of LOQC architectures.
\end{abstract}

\maketitle


\section{Introduction}
\begin{figure*}
    \centering
    \includegraphics[width=\textwidth]{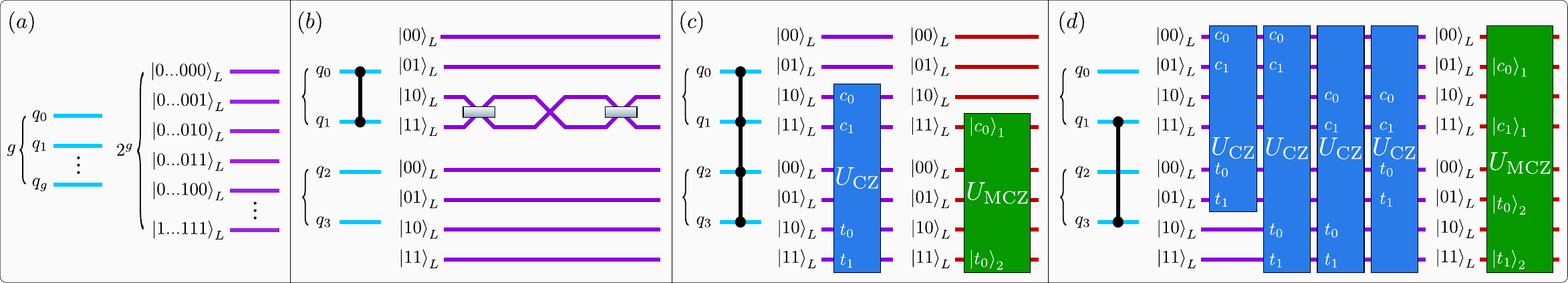}
    \caption{A schematic description of a qudit compression scheme. Here the blue wires represent the encoding of qubits onto qudits, with the brackets to the left indicating the qubits encoded onto a qudit and the purple and red wires represent the corresponding optical circuit implementation using existing and our proposed circuit compression schemes respectively. (a) Represents $g$ qubits mapped to the same qudit comprising of $2^{g}$ spatial modes. (b) A non-local CZ gate between two qubits mapped to the same qudit is transformed into a local gate. (c) Shows the optical implementation of $4$ qubit CCCZ gate where $2$ qubits are encoded onto each qudit using circuit compression. (d) Shows the optical implementation of a compressed circuit where two control qubits are removed from the multi-level CZ gate (one from each qudit).}
    \label{fig:QLOQ_description}
\end{figure*}
Linear optical quantum computation (LOQC) has emerged as a strong candidate for scalable and fault-tolerant quantum computation~\cite{AC99, KLM01, RB01, OPW03,WRR05, KMN07, BBD09, AA13, WQD19, WSL20, SDS24, RM25, LBK25, BAV21, LBK25,psi25}, in which a single photon can encode a qubit or a $d$-dimensional qudit \cite{PF21, MLS24, CHZ22, IJA19, LBB09, HXG18, EMF17, LWW20, ZXQ25, MEH16} by occupying one of $2$ or $d$ spatial modes respectively~\cite{RZB94,WPB16}. LOQC promises successful implementation of single qubit and qudit quantum gates with near unit fidelity~\cite{PBJ21, CRK24, M22} using only passive linear optical elements like phase plates and beam splitters (BS) together with single photon sources and detectors. 
Integrated photonic implementations can now incorporate on-chip detectors, switches, delays and multiplexing ~\cite{KBP24, NAM24, ALC19} establishing LOQC as a viable platform for practical and scalable quantum computation~\cite{R17}.

Despite these advantages, the scalability of LOQC is fundamentally limited by probabilistic nature of two-qubit non-local entangling gates between photons. Even in the best case, such two-qubit entangling gates only succeed with probability $\frac{1}{9}$~\cite{RLB02} which involves post-selection. Furthermore, non-local entangling gates in LOQC also require additional ancillary photons and spatial modes increasing the cost of resources required.
Consequently, for achieving scalable LOQC it is imperative to either reduce the number of non-local entangling gates or lower their resource cost.

Qudit circuit compression schemes~\cite{GAF23,LSG24,NKF24,KNF25,NZB25} have been developed to address these challenges which exploit the rich structure of $d$-dimensional qudits for efficient quantum information processing and computation~\cite{LBB09,C14,WCA15,KNX20,WHS20,NKF24}. 
Qudit circuit compression is a circuit compilation technique that maps a given $N$-qubit unitary circuit onto a $M$-qudit unitary circuit ($M < N$) with the goal of converting the maximum number of non-local multi-qubit entangling gates into local qudit gates which can be implemented deterministically.
In such schemes $N$ photonic qubits are mapped onto $M$ qudits (using $M$ single photons), where the $i$th qudit encodes $g_i$ number of qubits using $2^{g_i}$ spatial modes such that $\sum_{i=1}^{M}g_i=N$ as shown in Fig.~\ref{fig:QLOQ_description}(a). 
This approach offers the following advantages: (i) previously non-local entangling gates across multiple qubits mapped to the same qudit are converted to local gates, which can be implemented with unit probability 
(see Fig.~\ref{fig:QLOQ_description}(b)) and (ii) the non-local entangling operations on all the $g_i$ qubits encoded on the first qudit and all the $g_j$ qubits encoded on a second qudit can be implemented using a single non-local entangling gate as opposed to implementing $2(g_i+g_j)-3$ non-local gates without compression
(see Fig.~\ref{fig:QLOQ_description}(c)). From the qudit perspective, this single non-local entangling gate constitutes a two-level CZ gate that conditionally triggers when both the qudits occupy the highest computational basis state $\ket{d - 1}$. An optical implementation of such a quantum gate has been shown in Ref.~\cite{LIU26} using orbital angular momentum degrees of freedom of photons.

A critical limitation of the qudit compression schemes arises when only a subset of the qubits encoded onto multiple qudits are required to act as conditions which trigger the non-local entangling gate. For these selective multi-control entangling gates, removal of each control (or target) qubit from the encoded qudits effectively doubles the number of non-local entangling gates required for the implementation, thereby nullifying the advantage of the circuit compression schemes.
This issue mainly arises due to the non-existence of multi-level entangling gates between qudits in which multiple and arbitrary levels across the two qudits can be selectively chosen to trigger the gate.

In this paper, we demonstrate the existence of multi-level control-Z (CZ) gates for LOQC based on spatial modes by explicitly providing two linear optical schemes. Unlike the two-level CZ implementations on qudits (known to be sub-optimal for a qudit architecture) which restrict the phase flip to a single pair of modes (typically the highest levels $|d-1\rangle_1 |d-1\rangle_2$)~\cite{LIU26}, multi-level CZ gates impart a conditional phase flip whenever the qudits occupy any combination of modes from a pre-defined set.

Our first scheme establishes the physical realizability of such a multi-level entangling operation with a success probability of $1/8$, which is independent of the number of qubits involved in the operation. However, this implementation requires \textit{a priori} knowledge about the initial qudit states. Consequently, it cannot be regarded as a standard quantum gate, but rather as an operation tailored to arbitrary yet known input quantum states.

On the other hand, the second scheme, which is derived from the first, realizes the same operation independent of the input state with a success probability of $\frac{1}{2} \left(\frac{1}{8}\right)^{2^{r_1}+2^{r_2}}$, where $r_1$ and $r_2$ are the number of qubits to be removed as controls from each qudit encoding. 
Both of our proposed schemes rely on the use of ancillary photons and Bell state measurements (BSM) coupled with post-selection~\cite{G11, P11, WHF16}. 

Coupled with circuit compression schemes, our proposed realizations provide the following circuit compilation advantages: (i) enables efficient qudit circuit compression by allowing selective removal of $r_1$ and $r_2$ number of qubits encoded onto two qudits for multi-level entangling gates such that they can be physically realized using at most only $\mathcal{O}(2^{r_1}+2^{r_2})$ non-local entangling gates which offers substantially better scaling as opposed to $\mathcal{O}(2^{r_1+r_2})$ gates in standard circuit compression schemes and, (ii) any desired multi-level entangling operation involving an arbitrary number of qubits across the two qudits can be achieved at a constant success probability of $1/8$ if the input state is known a priori.

Moreover, both the schemes are implementable on integrated photonic architecture, and are compatible with the qudit based architecture of quantum computation.
To demonstrate the usefulness of our schemes, we also realize the quantum full adder (QFA) circuit using them. We show that our schemes can significantly reduce the total number of non-local entangling gates required when used in conjunction with qudit circuit compression schemes.
Overall, our realizations address a key limitation of existing circuit compression schemes and offer better scalability for LOQC.

The paper is organized as follows. In Sec.~\ref{sec:multi_gates} we introduce the concept of multi-level entangling gates in LOQC following the notion of multi-qudit entangling gates. In Sec.~\ref{subsec:state_dependent} we demonstrate our scheme when the input quantum state can be arbitrary but known. In Sec.~\ref{subsec:state_independent} we generalize our scheme for arbitrary and unknown input quantum states. In Sec.~\ref{sec:application} we illustrate an application of our realizations for the quantum full adder.
In Sec.~\ref{sec:conc} we conclude our results by discussing their advantages and disadvantages, implementation for circuit compilation and other applications. 


\section{Multi-level entangling gates}
\label{sec:multi_gates}

In qudit circuit compression schemes, a single photon having access to $d = 2^{g}$ spatial modes encodes a single qudit (alternatively it can be seen that $g$ qubits are mapped to a single qudit). This qubit-to-qudit mapping ensures that a non-local entangling gate between any of the two qubits is converted into a local gate (as shown in Fig.~\ref{fig:QLOQ_description}(b)). Moreover, if multiple qubits are encoded onto two qudits, then a controlled operation comprising of all the qubits encoded on two qudits is converted into a single non-local entangling gate, hereafter termed as a two-level CZ gate, as it conditionally triggers when the physical state of the two qudits are $\ket{d-1}_1$ and $\ket{d-1}_2$, respectively, corresponding to the logical states $\ket{11\ldots1}_1$ and $\ket{11\ldots1}_2$ (an example for $d=4$ can be seen in Fig.~\ref{fig:QLOQ_description}(c)). This two-level CZ gate can be implemented probabilistically in LOQC (upto local unitaries)~\cite{RLB02}.  It applies a control phase flip operation when both the qudits are in the state $\ket{d - 1}$. Mathematically, it is represented as a unitary matrix $U_{\text{CZ}}$ such that
\begin{equation}
    U_{\text{CZ}} = \mathds{1} - 2 \ket{d - 1}\!\bra{d - 1}_1\otimes \ket{d - 1}\!\bra{d - 1}_2,
    \label{eq:cz_gate_qubits}
\end{equation}
where $\mathds{1}$ denotes a $d^2\times d^2$ identity matrix and the subscript on the states denotes the qudit to which it belongs. 
Operationally, an optical implementation of this gate is equivalent to choosing the optical modes corresponding to states $\ket{d - 1}$ of first and second qudits as the conditions that triggers the CZ gate. Hereafter we term these optical modes as trigger modes (see Fig.~\ref{fig:QLOQ_description}(c)). 

While the aforementioned two-level CZ gate is a straightforward extension of the standard two-qubit CZ gate with only a single trigger mode in each qudit, its generalization to the construction of a single quantum gate in which there are multiple trigger modes of the first and second qudits, is non-trivial. 
We provide an optical construction of the same using linear optical elements as follows.

Henceforth, we will focus on constructing a multi-level analog of the two-level CZ gate with multiple trigger modes. We term this analogous gate as a multi-level entangling gate (due to the fact that multiple and arbitrary trigger modes can be chosen).

We begin by elucidating the action of the multi-level entangling gate on two photons, each having access to $d = 2^{g}$ spatial modes which can jointly encode two qudits. Let us also define the set of indices corresponding to the trigger modes (which can be chosen arbitrarily but are in general always known a priori to the application of the gate) in the first and second qudit by $\mathcal{C}_1 = \lbrace c_0,c_1, \ldots, c_{k_1-1}\rbrace$ and $\mathcal{C}_2 = \lbrace t_0, t_1, \ldots, t_{k_2-1}\rbrace$ respectively,
where $k_1, k_2 < d$ such that each pair of indices $\left( c_i,t_j\right) \in \mathcal{C}_1 \times \mathcal{C}_2$ identifies the states $\ket{c_i}_1$ and $\ket{t_j}_2$ as the trigger modes for the application of the quantum gate. The action of the multi-level entangling gate is then given by the unitary matrix 
\begin{equation}
    U_{\text{MCZ}} = \mathds{1}-2\sum_{m \in \mathcal{C}_1}\sum_{n \in \mathcal{C}_2}\ket{m}\bra{m}_1 \otimes \ket{n}\bra{n}_2,
\label{eq:gate_equation}
\end{equation}
where its action on the set of states $\lbrace \ket{m}_1\ket{n}_2~|~  (m,n)\in \mathcal{C}_1 \times \mathcal{C}_2\rbrace$ is to apply a phase flip, mapping each state to $-\ket{m}_1\ket{n}_2$ respectively, while the set of states $\lbrace \ket{m}_1\ket{n}_2~ |~  m \notin \mathcal{C}_1 ~\text{or}~ n\notin \mathcal{C}_2~\text{(or both)} \rbrace$ remain unchanged.


\begin{figure}
    \centering    \includegraphics[width=1\linewidth]{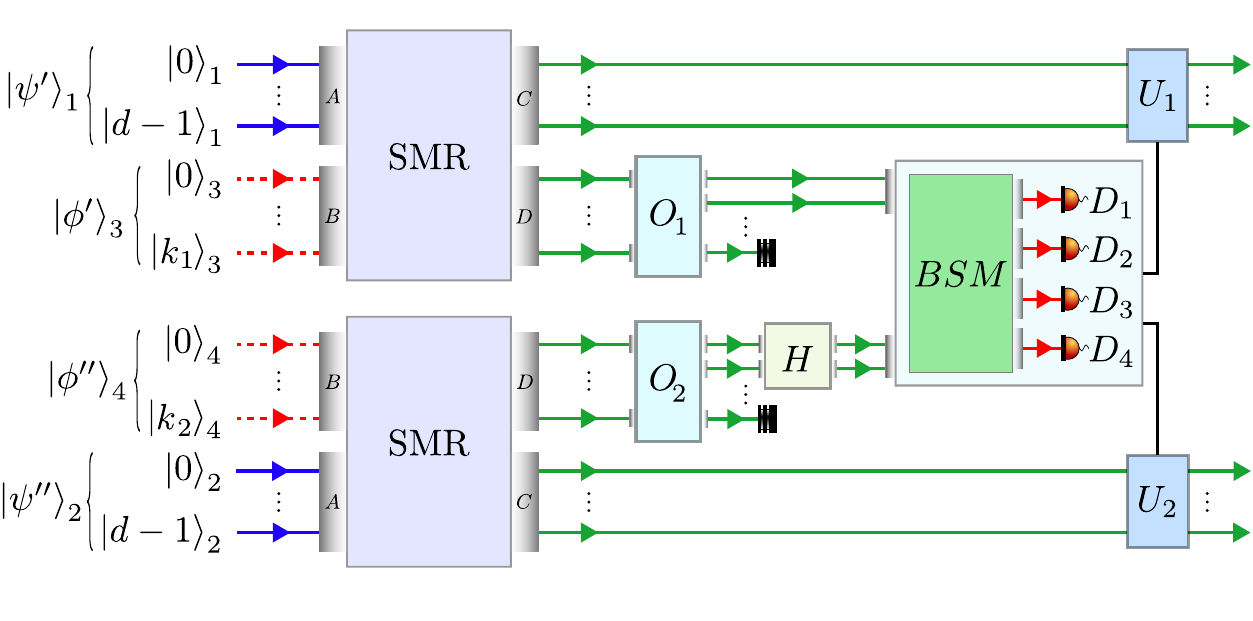}
    \caption{Schematic of the state dependent multi-level CZ operation. Two input-ancilla pairs in (input in blue, ancilla in red) serve as inputs to the two SMRs. The SMRs execute a partial swap operation on the defined trigger sets. Only the case in which one photon exits the port of the two SMRs is post-selected. The ancilla qudits $3$ and $4$ are then fed into unitaries $O_1$ and $O_2$ which, based on the knowledge about input qudits, map them to a two-dimensional subspace while the remaining $k_i-1$ modes are unoccupied. Finally, a Hadamard gate is applied to photon $4$ before a BSM is implemented on the qudits $3$ and $4$. Depending on the outcomes, the local unitaries $U_1$ and $U_2$ are then implemented, finally realizing the desired operation.}
    \label{fig:Gate_and_SMR}
\end{figure}

\section{Linear optical schemes for multi-level entangling operations}
\label{sec:linear_schemes}

In this section we demonstrate our linear optical schemes for implementing multi-level entangling operations between two input qudits. We first elucidate a state dependent scheme which subsequently forms the basis for the state independent version.

\subsection{State dependent scheme}
\label{subsec:state_dependent}

To realize the operation physically, we proceed in four stages. First, a pair of ancillary qudits are prepared to facilitate the implementation of the multi-level CZ operation, along with the two input qudits on which the operation needs to be implemented. Second, on each pair of the input qudits and an ancillary qudit we implement an operation to effectively flag whether the state of the input qudit corresponds to a trigger mode without collapsing its state. Third, a phase flip is imparted to the joint state of the input qudit and the ancillary qudits  whenever the state of the input qudit is flagged as a trigger mode. Finally, a BSM is implemented on the ancillary qudits, effectively realizing the multi-level CZ operation on the input qudits upto local unitaries.

Consider two photonic qudits, each encoded in $d$ spatial modes. Let
$\mathcal{B}_1, \mathcal{B}_2 = \lbrace\ket{0},\ket{1},...,\ket{d-1}\rbrace$ denote orthonormal bases for the Hilbert spaces $\mathcal{H}_1$ and $\mathcal{H}_2$ associated with the spatial modes of the two photons, such that
\begin{equation}
    \ket{\psi'}_1 = \sum_{m = 0}^{d - 1}\beta_m \ket{m}_1, \quad \ket{\psi''}_2 = \sum_{n = 0}^{d - 1}\gamma_n \ket{n}_2,
\end{equation}
are some arbitrary states over the two qudits where $\beta_m, \gamma_n \in \mathbb{C} ~\forall m,n$ and $\sum_{m = 0}^{d - 1} \abs{\beta_m}^2 = \sum_{n = 0}^{d - 1} \abs{\gamma_n}^2 = 1$. Any arbitrary product state over these two qudits can then be written as $\ket{\psi}_{12} = \ket{\psi'}_1\ket{\psi''}_2$. 
We introduce two ancillary qudits $\ket{\phi'}_3$ and $\ket{\phi''}_4$ with Hilbert space dimensions $k_1 + 1$ and $k_2 + 1$ respectively, whose systems are labeled by $3$ and $4$ (see Fig.~\ref{fig:Gate_and_SMR}). The systems corresponding to the ancillary qudits is chosen such that for every trigger mode $\ket{c_i}_1$ in the first qudit, we identify a corresponding mode $\ket{i}_3$ for $i \in \lbrace 0,1,\ldots,k_1-1 \rbrace$. In addition to these $k_1$ modes, we include an additional independent mode $\ket{k_1}_3$ such that $\lbrace \ket{i}_3\rbrace_{i = 0}^{k_1}$ forms an orthonormal basis.
Similarly, the system corresponding to the second ancillary qudit is chosen such that for every trigger mode $\ket{t_j}_2$ in the second qudit, we identify a corresponding mode $\ket{j}_4$ for $j \in \lbrace 0,1,\ldots,k_2-1 \rbrace$ and an additional independent mode mode $\ket{k_2}_4$, such that $\lbrace \ket{j}_4\rbrace_{j = 0}^{k_2}$ forms an orthonormal basis.

The two ancillary qudits are initialized in the state
\begin{equation}
\ket{\phi'}_3 = \frac{1}{\sqrt{2}}\left(\ket{\xi_1}_3 + \ket{k_1}_3 \right), \quad\ket{\phi''}_4 = \frac{1}{\sqrt{2}}\left(\ket{\xi_2}_4 + \ket{k_2}_4 \right),
\label{eq:ancillas}
\end{equation}
where
\begin{equation}
\ket{\xi_1}_3 = \frac{1}{\sqrt{P_1}} \sum_{i |c_i \in \mathcal{C}_1} \beta_{c_i} \ket{i}_3, \quad \ket{\xi_2}_4 = \frac{1}{\sqrt{P_2}} \sum_{j|t_j \in \mathcal{C}_2} \gamma_{t_j} \ket{j}_4,
\label{eq:normalizied_states}
\end{equation}
and $P_1 =\sum_{c_i \in \mathcal{C}_1}\abs{\beta_{c_i}}^2$ and $P_2 =\sum_{t_j \in \mathcal{C}_2}\abs{\gamma_{t_j}}^2$.

Following Eq.~\eqref{eq:ancillas} and \eqref{eq:normalizied_states} it can be seen, for $k_1,k_2>1$, a priori knowledge about the initial quantum states $\ket{\psi'}_1$ and $\ket{\psi''}_2$ is required to prepare the ancilla qudits. As a consequence, the physical realization we propose does not yet constitute a proper quantum gate in the traditional sense, where a single fixed physical realization is required to act identically on arbitrary and unknown input states. We provide solutions to deal with this issue later.
On the other hand, for $k_1,k_2=1$, our proposed realization functions as a valid quantum gate for arbitrary and unknown input qudit states.

\begin{figure}
    \centering    \includegraphics[width=1\linewidth]{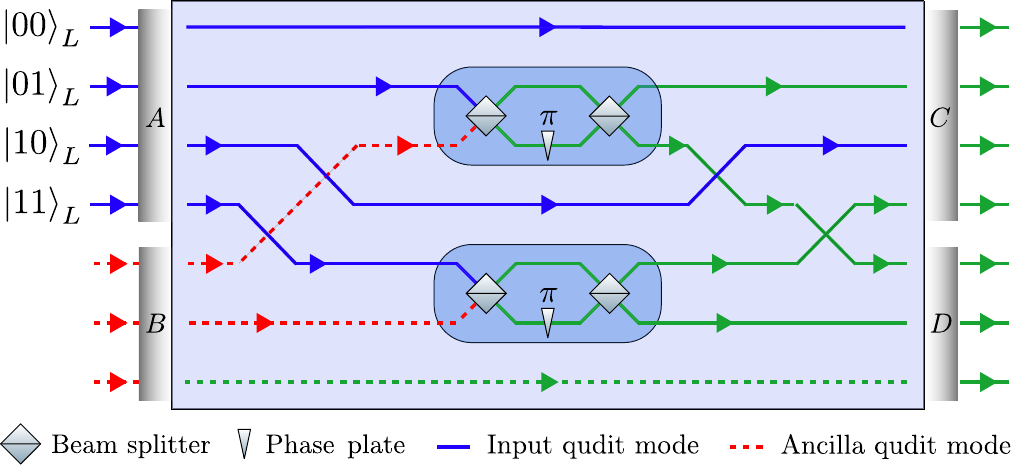}
    \caption{Optical realization of an SMR for $4$-dimensional input qudit with $\ket{01}_L$ and $\ket{11}_L$ as the trigger modes, as in Fig.~\ref{fig:QLOQ_description}(d). The phase difference across the two arms of Mach-Zehnder interferometers (dark blue boxes), are set to $\pi$ to facilitate the swap between the desired input and ancilla modes.
    \label{fig:4D_SMR}}
\end{figure}

After initialization, the qudits $1$ and $2$ are paired with the ancillary qudits $3$ and $4$ respectively, using a selective mode router (SMR) as shown in Fig.~\ref{fig:Gate_and_SMR}. The SMR features two input ports, labeled $A$ and $B$, and two output ports labeled $C$ and $D$. The qudits $1$ and $2$ are fed into the input ports $A$, while the ancilla qudits $3$ and $4$ are fed into the ports $B$ of the two SMRs, respectively.
The action of the SMR on any two states $\ket{x}_A \ket{y}_B$, $x, y \in \lbrace0, 1,\ldots, d - 1 \rbrace$, input in ports $A$ and $B$ for a set of trigger mode indices $\mathcal{C}_i$, $i\in \lbrace 1, 2 \rbrace$ is given by
\begin{equation}
\begin{aligned}
\mathrm{SMR}
\!\left( \ket{x}_A  \ket{y}_B \right)
=
\begin{cases}
\ket{x}_C  \ket{y}_D,
& x \notin \mathcal{C}_i,\ y \notin \mathcal{C}_i \\[4pt]
\ket{x,y}_C  \ket{\mathrm{vac}}_D,
& x \notin \mathcal{C}_i,\ y \in \mathcal{C}_i \\[4pt]
\ket{\mathrm{vac}}_C \ket{x,y}_D,
& x \in \mathcal{C}_i,\ y \notin \mathcal{C}_i \\[4pt]
\ket{y}_C \ket{x}_D,
& x \in \mathcal{C}_i,\ y \in \mathcal{C}_i
\end{cases}
\text{     ,}
\end{aligned}
\label{eq:smr_operation}
\end{equation}
where $\ket{\text{vac}}$ represents the vacuum state (no photon) and $\ket{x,y}_C$ denotes a two-photon state in the output port $C$ (same for port $D$). As can be seen, the SMR acts like a (partial) swap operation, exchanging the output ports if both of the input states correspond to trigger modes. In other cases it either acts like an identity transformation (when both $x, y\notin \mathcal{C}_i$) or we observe photon bunching (when either $x$ or $y\in \mathcal{C}_i$). 

A general optical implementation of the SMR in given in the Appendix \ref{appendix:SMR}. The SMR consists solely of an array of Mach-Zehnder interferometers, which are configured to swap the trigger modes of the input qudit and the ancillary qudit. As an example Fig.~\ref{fig:4D_SMR} demonstrates the action of an SMR for the set of trigger modes in the first qudit $\mathcal{C}_1 = \lbrace 1,3 \rbrace$ (corresponding to the example in Fig.~\ref{fig:QLOQ_description}(d)). The Mach-Zehnder interferometers (with the phase difference set to $\pi$) facilitate the swap operation between the trigger modes of the input qudit and the ancillary qudit. The remaining modes of the input qudit and the additional mode of the ancillary qudit pass as such.

After the action of the two SMRs we post-select on those cases in which each output port emits only a single photon. The bunching events constitute a failure, leading to a probabilistic nature of our realization. Following Eq.~\eqref{eq:smr_operation}, the probability of a successful SMR operation (i.e. no bunching of photons) is $1/2$ (for more details see Appendix \ref{appendix:SMR}).

Following the action of SMR, we implement unitary transformations $O_1$ and $O_2$ independently on the ancillary systems $3$ and $4$ which map them onto a two dimensional subspace. This mapping can be implemented deterministically (following Reck and Clements multiport interferometer constructions~\cite{RZB94, WPB16}) conditioned on the fact that the initial quantum state is known (for more details see the Appendix \ref{appendix:unitaries}). Following the action of $O_1$ and $O_2$, the states of the input-ancilla pairs are given as 
\begin{subequations}
\begin{equation}
    \ket{\psi'}_{1} \ket{\phi'}_{3} = \sum_{m \notin \mathcal{C}_1} \beta_m \ket{m}_{1}\ket{0}_{3} + \sum_{m \in \mathcal{C}_1} \beta_m \ket{m}_{1}\ket{1}_{3},
\end{equation}
\begin{equation}
    \ket{\psi''}_{2} \ket{\phi''}_{4} = \sum_{n \notin \mathcal{C}_2} \gamma_n \ket{n}_{2}\ket{0}_{4} + \sum_{n \in \mathcal{C}_2} \gamma_n \ket{n}_{2}\ket{1}_{4},
\end{equation}
\label{eq:state_independent_requirement}
\end{subequations}

which constitutes an input-ancilla entangled state. Furthermore, it should be noted that when $k_1,k_2 = 1$, $O_1$ and $O_2$ act as an identity transformation. 

Afterwards, we implement a Hadamard gate on the ancilla qudit $4$. Following this, the final joint state of all the $4$ qudits is
\begin{equation}
\begin{aligned}
\ket{\Psi}
&= U_{\text{MCZ}}\ket{\psi}_{12} \otimes \ket{\phi^+}_{34}
    \\
&\quad
   + \left[(U_1 \otimes \mathds{1})U_{\text{MCZ}}\ket{\psi}_{12}\right] \otimes \ket{\phi^-}_{34}\\
&\quad
    + \left[(\mathds{1} \otimes U_2) U_{\text{MCZ}}\ket{\psi}_{12}\right] \otimes \ket{\psi^+}_{34}\\
&\quad   
   + \left[(U_1  \otimes U_2) U_{\text{MCZ}}\ket{\psi}_{12}\right] \otimes \ket{\psi^-}_{34},
\end{aligned}
\end{equation}

where $\lbrace \ket{\phi^+}_{34}, \ket{\phi^-}_{34}, \ket{\psi^+}_{34}, \ket{\psi^-}_{34} \rbrace$ are the four Bell states and $U_1$ and $U_2$ are some local unitaries that act on the qudits $1$ and $2$ respectively. The desired operation $U_{\text{MCZ}}$ on an initial state $\ket{\psi}_{12}$ can then be realized by performing a BSM on the joint state of the qudits $3$ and $4$ and implementing the local unitary operations $U_1$ and $U_2$ based on the observed outcomes.

The success of the operation is heralded by a four-fold coincidence detection. Since each SMR succeeds with probability $1/2$, the overall success probability is $1/4$. Furthermore, employing linear optics for the BSM restricts us to distinguishing only two of the four Bell states with probability $1/2$. Consequently, the final probability for a successful linear optical implementation of $U_{\text{MCZ}}$ is $1/8$. 

We also numerically realize our state dependent multi-level entangling operation~\cite{multilevel} on the open source platform Perceval~\cite{HFG23} for linear optical quantum computation.


\begin{figure}
    \centering
    \includegraphics[width=1\linewidth]{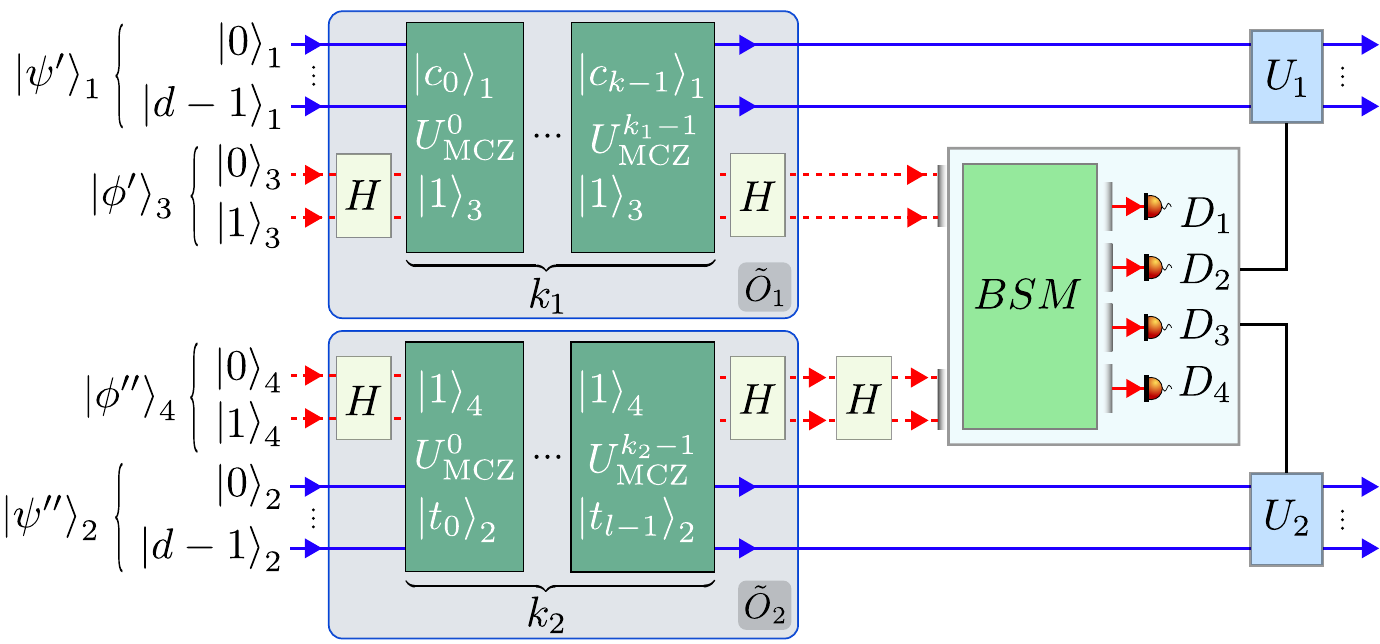}
    \caption{Schematic of the state independent multi-level CZ gate. For $k_1$ ($k_2$) number of trigger modes in qubit $1$ ($2$), a sequence of $k_1$ ($k_2$) two-level CZ gates between the input qubit $1$ and the ancilla qubit $3$ (along with two Hadamard gates), selectively flips the ancilla qubit whenever input qudit is in the trigger state. The green boxes show two-level CZ gates, each triggered by $\ket{c_i}_1$ and $\ket{1}_3$ ($\ket{t_j}_2$ and $\ket{1}_4$). The grey boxes constitute the sequence of operations $\Tilde{O}_1$ and $\Tilde{O}_2$ respectively.  A BSM is then performed on the ancilla qubits followed by the required unitary operations.}
    \label{fig:state_independent}
\end{figure}

\subsection{State independent scheme}
\label{subsec:state_independent}

In order to have a scheme that does not require a priori knowledge of the input qudits $1$ and $2$ (following the definition of standard quantum gates), it is necessary for the input qudit and the ancilla be entangled before the BSM. Specifically, we require them to be in the state given by Eq.~\eqref{eq:state_independent_requirement} (see Appendix \ref{appendix:state_independent} for more details). To generate the requisite entangled state, we now consider qubit ancillas (as opposed to qudits in the state dependent scheme). The core idea is to sequentially apply the multi-level CZ operation on the input qudit and qubit ancilla pair each consisting of single trigger modes. Note that for single trigger modes, our realization of the multi-level CZ gate is state independent, constituting a proper two-level CZ gate in the traditional sense.

We begin with the ancilla qudits initialized in the state $\ket{0}_{i + 2}$ ($i\in\lbrace1, 2\rbrace$). Afterwards we implement a composite unitary operation (for more details see Appendix \ref{appendix:state_independent}) 
\begin{equation}
    \Tilde{O}_i = \left(\mathds{1} \otimes H\right)\left(\prod_{s = 0}^{k_i-1} U_{\text{MCZ}}^{s}\right) \left(\mathds{1} \otimes H\right),
\label{eq:state_independent_operator}
\end{equation}
where $U_{\text{MCZ}}^{s}$ denotes the $s$th two-level CZ gate between the qudits $i$ and $i + 2$ realized by our proposed construction, such that the two trigger modes correspond to the indices $s$ and $1$ respectively and $H$ is a Hadamard gate implemented on the ancilla qubits. Here, it should be noted that each instance of $U_{\text{MCZ}}^{s}$ is physically realized with the help of two ancilla qubits, amounting to a total of $2(k_1+k_2)$ ancilla qubits for the operation $\tilde{O}_i$. See Fig.~\ref{fig:state_independent} for a schematic representation. 

Following the application of $\Tilde{O}_i$, we proceed similarly as before, implementing a BSM on the ancillary qubits and applying the requisite local unitary operation based on the outcomes. We finally realize the operation $U_{\text{MCZ}}$ on the state $\ket{\psi}_{12}$ with a success probability of $\frac{1}{2} \left(\frac{1}{8}\right)^{k_1+k_2}$, following the fact that each two-level CZ gate has a success probability of $1/8$, and we use $k_1+k_2$ such gates, followed by a BSM that succeeds with probability $1/2$.

Following our state-independent realization we find, that for circuit compression schemes which require the removal of $r_1$ and $r_2$ number of qubits (where $k_1 = 2^{r_1}$ and $k_2 = 2^{r_2}$) as controls across the two qudits respectively, $2^{r_1}+2^{r_2}$ number of probabilistic entangling gates are required to realize the operation (following the fact that removal of $r$ qubits as control from one qudit requires $2^r$ modes to act as triggers in the corresponding qubit). This is in contrast with the current best bound of $2^{r_1+r_2}$ which is exponentially higher.

It should be noted that for sufficiently large values of $r_1$ and $r_2$, the state-independent scheme would still fail to serve the purpose of noisy-intermediate-scale-quantum devices unless circuit compression is applied optimally. Instead of directly translating a qubit based circuit into a qudit based circuit, the different
qubits need to be grouped and encoded onto qudits in a manner following the results of Ref.~\cite{LSG24}. We illustrate this point with an example in the next section.


\section{Application: Qudit circuit compression for the quantum full adder}
\label{sec:application}

\begin{figure*}
    \centering
    \includegraphics[scale = 0.4]{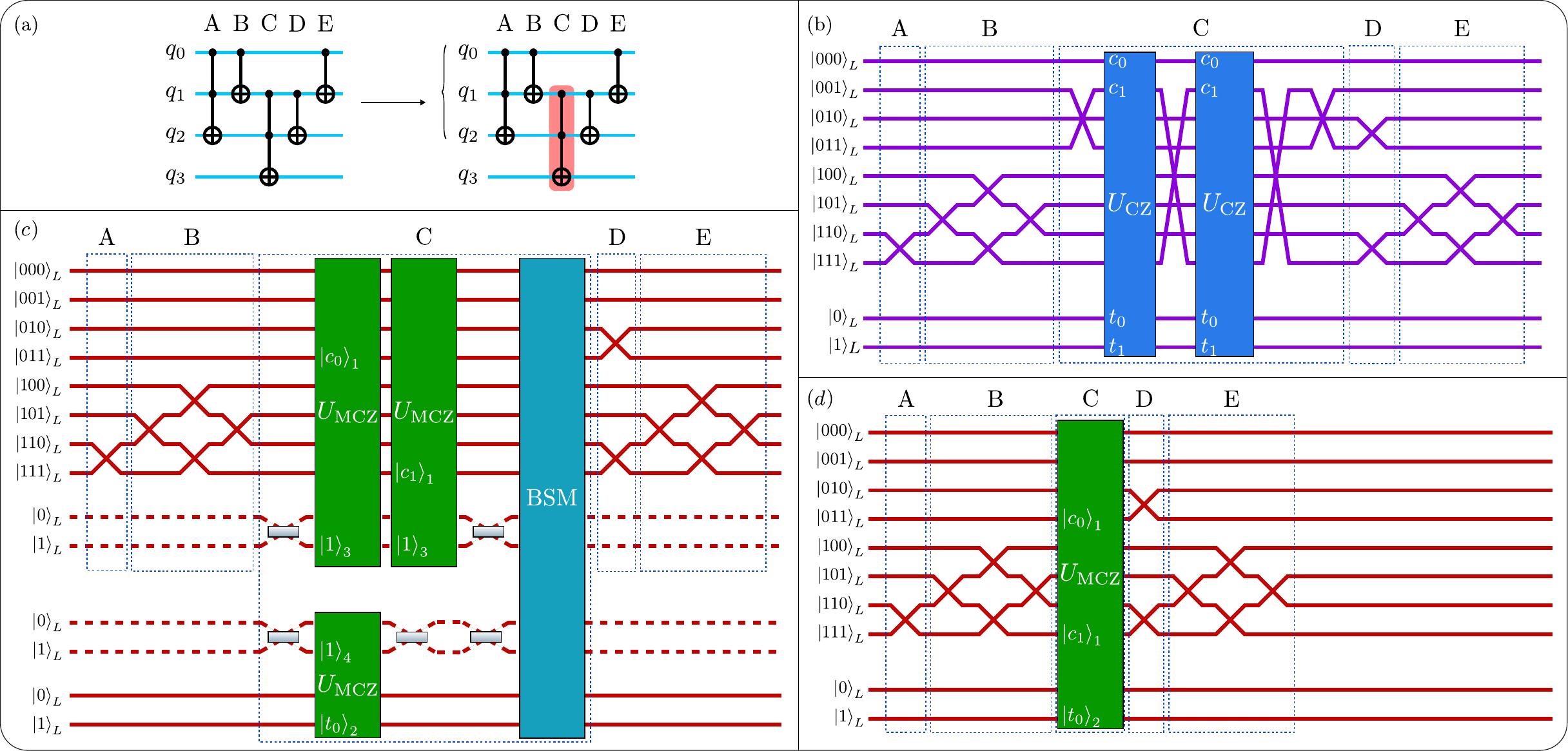}
    \caption{(a) A schematic of the QFA circuit using (b) currently available non-local CZ gates, (c) our proposed state-independent multi-level CZ gate and, (d) state-dependent multi-level CZ operation. The input qubits $q_0, q_1$, and $q_2$ are encoded onto a single qudit.
    (b) In qudit circuit compression schemes, the operation corresponding to C is a non-local CNOT gate that must be decomposed into two non-local entangling gates for a linear optical realization. (c) For qudit compression scheme, the same circuit would require three non-local entangling gates using our state-independent scheme, followed by the Bell state measurement on the ancilla qubits, and the feedforward corrections $U_1$ and $U_2$.
    (d) The same operation corresponding to C can be implemented using a single multi-level entangling operation via our state-dependent scheme. In order to prepare the ancilla qudits for the multi-level entangling operation they are first initialized in the same state as the input qudits and then subsequently subject to the local operations A and B.}
    \label{fig:QFA}
\end{figure*}

To demonstrate an application of our proposed realizations of multi-level CZ gates, we apply it to an optical realization of quantum full adder (QFA) circuit consisting of $4$ qubits. The circuit implements binary addition on three input qubits, representing the two summands and the input carry, to produce the arithmetic sum and the output carry, represented by the third and the fourth qubits, respectively.

A circuit for the same can be realized using three non-local CNOT (CX) gates and two Toffoli (CCX) gates as shown in Fig.~\ref{fig:QFA}(a). In LOQC, a CCX gate is typically decomposed into three CX gates, raising the total CX gate count in the realization to $9$. Furthermore, we note that a non-local CZ gate can be transformed into a non-local CX gate by simple local unitaries consisting only of BS.

Next, using the fact that a CX gate can only be successfully realized with a probability $1/9$ in LOQC, the total success probability of implementing the QFA circuit scales as $\left(1/9\right)^9$, which is vanishingly small. 
To mitigate this issue we look at qudit compression schemes to reduce the number of non-local entangling gates.

Let us consider that the qubits $q_0$, $q_1,$ and $q_2$ are mapped onto a single qudit with dimension $d=8$ as shown in Fig.~\ref{fig:QFA}(a). In this encoding, we see that all non-local CX gates are converted into local qudit gates except one. The operation corresponding to C in Fig.~\ref{fig:QFA} remains a non-local entangling gate. As can be seen this operation requires the removal of $q_0$ as the control qubit. Following standard circuit compression schemes \cite{LSG24, KNF25} an optical realization of the same would necessitate two CX gate operations such that the probability of successfully implementing the QFA circuit is $\left(1/9\right)^2$.

On the other hand, our proposed state dependent realization of multi-level CZ operation can implement this same circuit with a success probability of $1/8$. In order to initialize the ancillary qudits for the operation corresponding to C in Fig.~\ref{fig:QFA}, they are first initialized in the same state as the input qudits before the operation corresponding to A. Next, they are passed through the same set of local operations A and B as the input qudits to prepare them in the desired state. It is important to note here that the aforementioned way of initializing the ancilla qudits is deterministic because the operations A and B are local. If in case they were non-local entangling gates, then our aforementioned way of preparing the ancilla qudits cannot be applied. As such, our state dependent realization can only be realized for those multi-level entangling gates which are preceded by only local gates.

A state-independent realization of our proposed multi-level CZ gate can be employed without any assumptions but with a trade-off. It requires three instances of two-level CZ gates with a total success probability of $\frac{1}{2}(\frac{1}{8})^3$. As can be seen in Fig.~\ref{fig:QFA}(c), the two-level CZ gate is applied twice for two combination of triggers for the first input-ancilla pair (ancilla in dotted red) to generate the input-ancilla resource state. Moreover, it is applied once more for the second pair for the appropriate trigger pair.

Next, we explicitly demonstrate how to remove $q_0$ as the control qubit by employing our proposed realization of multi-level CZ gate. 

Consider the case when all three qubits $q_0,q_1,q_2$ act as control on the target qubit $q_3$. The trigger mode in that case would be the logical state $\ket{111}_L$ corresponding to the optical mode $\ket{7}_P$. As a result, removal of $q_0$ as a control qubit implies that the gate must triggered irrespective of the state of $q_0$. Hence, the logical states $\ket{011}_L$ and $\ket{111}_L$ must trigger the action of the gate. These states correspond to the optical modes $\ket{3}_P$ and $\ket{7}_P$. Following our proposal, these optical modes then constitute the trigger modes indexed by the set $\mathcal{C}_{1} = \{3,7\}$. 
On the other hand, trigger mode on $q_3$ is then indexed by the set $\mathcal{C}_{2} = \{1\}$. 

With the trigger sets identified and the ancilla qudits initialized in the desired state, we can implement the operation $U_{\text{MCZ}}$ which corresponds to implementing the operation corresponding to C in Fig.~\ref{fig:QFA}.

For the state-independent version as shown in Fig.~\ref{fig:QFA}(c), a Hadamard gate to both the ancilla qubits is applied first. Then, the optical modes $\ket{3}_P$ in the first input qudit and mode $\ket{1}$ in the first ancilla are used to trigger the first instance of the two-level CZ gate. It is followed by another instance which is triggered by the optical modes $\ket{7}_P$ in the first input qudit and $\ket{1}$ in the first ancilla qudit. The same is implemented for the qubit $q_3$ and the second ancilla qubit using one instance of the gate triggered by the optical mode $\ket{1}_P$ in the $q_3$ and $\ket{1}$ in the second ancilla. Then, a Hadamard gate is applied on both the ancilla qubits, followed by another Hadamard on second ancilla. Finally, a joint Bell state measurement is performed on the ancilla qubits followed by feedforward corrections $U_1$ and $U_2$.

Finally we also note that the qubit to qudit mapping used here is optimal for reducing the total number of non-local entangling gates. Any other mapping will always yield more than one non-local entangling gate.
For instance, $\lbrace q_0, q_1\rbrace$ can be encoded on the first qudit while $\lbrace q_1, q_2 \rbrace$ can be encoded on the second. In this case the entangling gates corresponding to B and E are now local and deterministic. Furthermore, the entangling gate corresponding to A can be tackled by our state-dependent scheme. However, the entangling gates corresponding to C and D cannot be tackled by our state-dependent scheme due to the requirement that the ancillary qudits need to be prepared by the actions of gates A and B, out of which, A already corresponds to a non-local multi-level entangling gate. As a consequence the entangling gates corresponding to C and D need to be decomposed into multiple two-level entangling gates respectively. Implementing the gate corresponding to C requires removal of one qubit as control from the first qudit (and none from the second qudit). Following the discussion in Sec.~\ref{subsec:state_independent} this requires $2^{r_1}+2^{r_2}=3$ non-local entangling gates. Similarly for D we require $2^{r_1}+2^{r_2}=4$ non-local entangling gates. This yields a total of $8$ non-local entangling gates to realize the QFA circuit instead of a single one following our optimal encoding.


\section{Conclusion}
\label{sec:conc}

In this paper, we have presented two linear optical realizations of multi-level CZ operation which are of significant importance enabling efficient qudit circuit compression schemes in which multiple (but in general not all) spatial modes of two qudits act as conditions to trigger the action of the non-local entangling gate. Since our schemes are native to a qudit based architecture, they are also of significant interest to high dimensional quantum computation \cite{MLS24, CHZ22, IJA19, LBB09, ZXQ25}.

Our first construction shows the physical realizability of the multi-level entangling operation with a successful implementation probability of $1/8$ independent of the dimension of the qudits. Importantly, in standard circuit compression schemes, this particular operation currently requires the sequential application of $2^{r_1 + r_2}$ probabilistic two-level CZ gates, where $r_1$ and $r_2$ number of qubits are to be removed as controls from the two qudits, respectively. In contrast, our scheme realizes the same operation with only a single multi-level CZ operation. 

However, this realization depends on a priori knowledge about the initial states of the two input qudits. As a consequence, this realization cannot be construed as a quantum gate, but rather as an operation on arbitrary but known input qudits.
While this is still of significant advantage in circuit compilation schemes, we also propose an alternate scheme which requires sequential application of our two-level gates and is independent of the initial quantum states. As such our second realization can be considered as a standard quantum gate. In order to remove $r_1$ and $r_2$ number of qubits as controls from the two qudits respectively, our second realization necessitates $2^{r_1} + 2^{r_2}$ number of probabilistic two-level CZ gates. This still performs exponentially better than the currently existing schemes. On the downside, the number of ancillary qubits required to implement the state independent scheme increases to $2(2^{r_1} + 2^{r_2})$.

Our proposed realizations of multi-qudit quantum operations provide a positive solution to a critical limitation of existing qudit circuit compression schemes by offering much better scalability in terms of number of probabilistic entangling gates. Our two realizations of the same multi-level entangling operations also demonstrate a fundamental trade-off between prior information about the input quantum states and the resources required to implement the gate.


\section{Acknowledgements}

The authors would like to acknowledge C.S. Yadav and R.P. Singh for valuable discussions. J.S. acknowledges financial support from the IIT Mandi seed grant project No. IITM/SG/JSN/168. A.T. acknowledges financial support from the Ministry of Education (MoE), Government of India, through the Half-Time Research Assistantship (HTRA) fellowship.



\appendix
\onecolumngrid

\section{Selective mode routers}
\label{appendix:SMR}

\begin{figure*}[h]
    \centering
    \includegraphics[scale = 0.6]{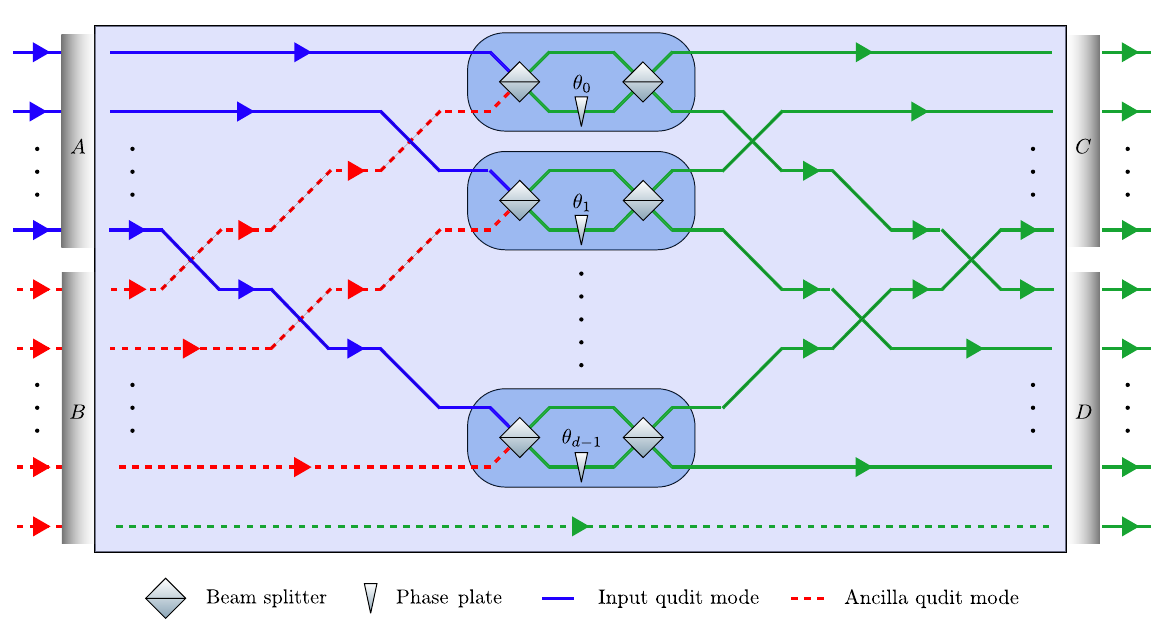}
    \caption{Linear optical realization of SMR. The two input ports $A$ and $B$ are fed the input qudit (represented by solid blue lines) and ancilla qudit (represented by dashed red lines) modes respectively. Only the trigger modes undergo the swap operation, which is facilitated by Mach-Zehnder interferometers (represented in blue boxes) parameterized by a phase shift $\theta_i$, $i \in \lbrace 0, 1, \ldots, d - 1 \rbrace$. For a $k$th mode which is chosen as a trigger mode, the parameter $\theta_k = \pi$, while for modes which are not triggers, $\theta_j = 0$. Ancilla is taken $(d+1)$-dimensional to allow flexibility in the choice of trigger modes. }
    \label{fig:SMR}
\end{figure*}

Let $\ket{\psi'}_1$ and $\ket{\psi''}_2$ represent the two qudits on which the multi-level control-Z (CZ) operation needs to be applied. 
We also define the set of indices corresponding to the trigger modes in the first and second qudit by $\mathcal{C}_1 = \lbrace c_0,c_1, \ldots, c_{k_1-1}\rbrace$ and $\mathcal{C}_2 = \lbrace t_0, t_1, \ldots, t_{k_2-1}\rbrace$ respectively, such that each pair of indices $\left( c_i,t_j\right) \in \mathcal{C}_1 \times \mathcal{C}_2$ identifies the states $\ket{c_i}_1$ and $\ket{t_j}_2$ as the trigger modes for the application of the quantum gate.
Consider two ancillary qudits $\ket{\phi'}_3$ and $\ket{\phi''}_4$ over $k_1 + 1$ and $k_2 + 1$ dimensional Hilbert space. Each of the ancillary qudits is formed by $k_i + 1$ spatial modes, $i\in \lbrace 1, 2\rbrace$. The initial quantum states corresponding to every system is given as
\begin{equation}
\begin{aligned}
\ket{\psi'}_1
&= \sum_{m=0}^{d-1} \beta_m \ket{m}_1, \text{ }
\ket{\psi''}_2 =
 \sum_{n=0}^{d-1} \gamma_n \ket{n}_2, \text{ }
\ket{\phi'}_3=
\frac{1}{\sqrt{2}}\left(\ket{\xi_1}_3 + \ket{k_1}_3 \right), \text{ }
\ket{\phi''}_4=
 \frac{1}{\sqrt{2}}\left(\ket{\xi_2}_4 + \ket{k_2}_4 \right),
\end{aligned}
\label{eq:definitions}
\end{equation}
where, 
\begin{equation}
\ket{\xi_{1}}_3 = \frac{1}{\sqrt{P_1}} \sum_{i|c_i \in \mathcal{C}_1} \beta_{c_i} \ket{i}_3, \quad \ket{\xi_{2}}_4 = \frac{1}{\sqrt{P_2}} \sum_{j|t_j \in \mathcal{C}_2} \gamma_{t_j} \ket{j}_4,
\label{eq:ancillas1}
\end{equation}

where $P_1 =\sum_{c_i \in \mathcal{C}_1}\abs{\beta_{c_i}}^2$ and $P_2 =\sum_{t_j \in \mathcal{C}_2}\abs{\gamma_{t_j}}^2$ are the corresponding normalization constants. The joint state of the input qudit $1$ and the ancilla qudit $3$ can then be written as
\begin{equation}
    \ket{\psi'}_1 \ket{\phi'}_3 = \frac{1}{\sqrt{2}}\sum_{m=0}^{d-1}  
    \left( \beta_{m} \ket{m}_1 \ket{\xi_1}_3
    + \beta_{m} \ket{m}_1 \ket{k_1}_3 \right).
\end{equation}

Similarly, we also have
\begin{equation}
    \ket{\psi''}_2 \ket{\phi''}_4 = \frac{1}{\sqrt{2}} \sum_{n=0}^{d-1}
    \left( \gamma_{n} \ket{n}_2 \ket{\xi_{2}}_4
    + \gamma_{n} \ket{n}_2 \ket{k_2}_4 \right),
\end{equation}

The selective mode routers SMR$_1$ and SMR$_2$ each act on the joint states $\ket{\psi'}_1 \ket{\phi'}_3$ and $\ket{\psi''}_2 \ket{\phi''}_4$ respectively. The SMR features two input ports, labeled $A$ and $B$, and two output ports labeled $C$ and $D$. The qudits $1$ and $2$ are fed into the input ports $A$, while the ancilla qudits $3$ and $4$ are fed into the ports $B$ of the two SMRs, respectively.
The action of the SMR on any two states $\ket{x}_A \ket{y}_B$, $x, y \in \lbrace0, 1,\ldots, d - 1 \rbrace$, input in ports $A$ and $B$ for a set of trigger mode indices $\mathcal{C}_i$, $i\in \lbrace 1, 2 \rbrace$ is given by
\begin{equation}
\begin{aligned}
\mathrm{SMR}
\!\left( \ket{x}_A  \ket{y}_B \right)
=
\begin{cases}
\ket{x}_C  \ket{y}_D,
& x \notin \mathcal{C}_i,\ y \notin \mathcal{C}_i \\[4pt]
\ket{x,y}_C  \ket{\mathrm{vac}}_D,
& x \notin \mathcal{C}_i,\ y \in \mathcal{C}_i \\[4pt]
\ket{\mathrm{vac}}_C \ket{x,y}_D,
& x \in \mathcal{C}_i,\ y \notin \mathcal{C}_i \\[4pt]
\ket{y}_C \ket{x}_D,
& x \in \mathcal{C}_i,\ y \in \mathcal{C}_i
\end{cases}
\text{     ,}
\end{aligned}
\label{eq:smr_operation_appendix}
\end{equation}
where $\ket{\text{vac}}$ represents the vacuum state (no photon) and $\ket{x,y}_C$ denotes a two-photon state in the output port $C$ (same for port $D$).
In our realization we post select on the cases where each exit port of both the SMR emits only a single photon. In an experimental setting, this is achieved via a four-fold coincidence measurement. A schematic diagram of an SMR is given in Fig.~\ref{fig:SMR}.

Let us first consider the action of SMR$_1$ on the joint state $\ket{\psi'}_1\ket{\phi'}_3$ where the state $\ket{\psi'}_1$ is fed into the port $A$ and $\ket{\phi'}_3$ is fed into the port $B$.

The output state after the action of SMR$_1$ on $\ket{\psi'}_1\ket{\phi'}_3$ is then given as
\begin{equation}
    \begin{aligned}
        \mathrm{SMR}_{1}\left(\ket{\psi'}_1 \ket{\phi'}_3\right)
        &= \frac{1}{\sqrt{2}}\left[\mathrm{SMR}_{1}
    \left(
        \sum_{m=0}^{d-1} \beta_m \ket{m}_1 \ket{\xi_{1}}_3
    \right)
    +
    \mathrm{SMR}_{1}
    \left(
        \sum_{m=0}^{d-1} \beta_m \ket{m}_1 \ket{k_1}_3
    \right)\right]\\
    &= \frac{1}{\sqrt{2}} \left[ 
    \frac{1}{\sqrt{P_1}}\mathrm{SMR}_{1}
    \left(
        \underset{\substack{m=0 \\ i\mid c_i \in \mathcal{C}_1}}{\sum^{d-1}} \beta_m \beta_{c_i} \ket{m}_1 \ket{i}_3
    \right)
    +
    \mathrm{SMR}_{1}
    \left(
        \sum_{m=0}^{d-1} \beta_m \ket{m}_1 \ket{k_1}_3
    \right)\right]\\
    &=\frac{1}{\sqrt{2}}\left[ 
    \begin{aligned}
    \frac{1}{\sqrt{P_1}}
        \mathrm{SMR}_{1}
        &\left(
            \underset{\substack{m \notin \mathcal{C}_1 \\ i\mid c_i \in \mathcal{C}_1}}{\sum} \beta_m \beta_{c_i} \ket{m}_1
            \ket{i} +
            \underset{\substack{c_{i'} \in \mathcal{C}_1 \\ i\mid c_i \in \mathcal{C}_1}}{\sum} \beta_{c_i'} \beta_{c_i} \ket{c_{i'}}_1
            \ket{i}_3
        \right)\\
        & + \mathrm{SMR}_{1}
         \left(
            \sum_{m \notin \mathcal{C}_1} \beta_m \ket{m}_1 \ket{k_1}_3
            +
            \sum_{c_{i'} \in \mathcal{C}_1} \beta_{c_{i'}} \ket{c_{i'}}_1 \ket{k_1}_3
        \right)
    \end{aligned}
    \right]\\
    &= \frac{1}{\sqrt{2}}\left[
    \begin{aligned}
        & \frac{1}{\sqrt{P_1}} \text{SMR}_1 \left(\underset{\substack{m \notin \mathcal{C}_1 \\ i\mid c_i \in \mathcal{C}_1}}{\sum} \beta_m \beta_{c_i} \ket{m}_1
             \ket{i}\right) 
            + \frac{1}{\sqrt{P_1}} \text{SMR}_1 \left(\underset{\substack{c_{i'} \in \mathcal{C}_1 \\ i\mid c_i \in \mathcal{C}_1}}{\sum} \beta_{c_{i'}} \beta_{c_i}  \ket{c_{i'}}_1
             \ket{i}_3\right)\\
        &\qquad +\text{SMR}_1\left(\sum_{m \notin \mathcal{C}_1} \beta_m \ket{m}_1 \ket{k_1}_3 \right)
        + \text{SMR}_1 \left(\sum_{c_{i'} \in \mathcal{C}_1} \beta_{c_{i'}} \ket{c_{i'}}_1 \ket{k_1}_3 \right)
    \end{aligned}
    \right]\\
    &= \frac{1}{\sqrt{2}}
    \left[
    \begin{aligned}
    \frac{1}{\sqrt{P_1}}
        &\left(
            \underset{\substack{m \notin \mathcal{C}_1 \\ c_i \in \mathcal{C}_1}}{\sum}
            \beta_m \beta_{c_i}
            \ket{m,c_i}_1 \ket{\text{vac}}_3
            + \underset{\substack{c_{i} \in \mathcal{C}_1 \\ i'\mid c_{i'} \in \mathcal{C}_1}}{\sum}
            \beta_{c_{i'}} \beta_{c_i}
            \ket{c_{i}}_1 \ket{i'}_3 
        \right)\\
        &\qquad + \left(
            \sum_{m \notin \mathcal{C}_1} \beta_m \ket{m}_1 \ket{k_1}_3
            +
            \sum_{i'|c_{i'} \in \mathcal{C}_1} \beta_{c_{i'}} \ket{\text{vac}}_1 \ket{i',k_1}_3
        \right)
    \end{aligned}
    \right],
    \end{aligned}
    \label{eq:SMR1_action}
\end{equation}
where in the second equality we have used Eq.~\eqref{eq:ancillas1} and in the third equality we have used the fact that a sum over the indices $m$ can be decomposed into a sum over $m\in \mathcal{C}_1$ and $m \notin \mathcal{C}_1$. Furthermore, in the third equality for the case when $m \in \mathcal{C}_1$ we replace the indices $m = c_{i'}$. In the last equality we have used the action of the SMR given in Eq.~\eqref{eq:smr_operation_appendix} and note that $\mathrm{SMR}_1 \left( \ket{c_{i'}}_1\ket{i}_3\right)=\ket{c_{i}}_1\ket{i'}_3$ where the trigger mode corresponding to the index $i'$ (denoted by $\ket{c_{i'}}_1$) in the qudit $1$ is swapped with the mode $i$ in the ancilla qudit (denoted by $\ket{i}_3$).

Similarly we have 
\begin{equation}
    \text{SMR}_2 \left(\ket{\psi''}_2\ket{\phi''}_4\right) = \frac{1}{\sqrt{2}} \left[
    \begin{aligned}
        &\frac{1}{\sqrt{P_2}}\left(
            \underset{\substack{n \notin \mathcal{C}_2 \\ j\mid c_j \in \mathcal{C}_2}}{\sum}
            \gamma_n \gamma_{t_j}
            \ket{n,t_j}_2 \ket{\text{vac}}_4
            + \underset{\substack{t_{j} \in \mathcal{C}_2 \\ j'\mid c_{j'} \in \mathcal{C}_2}}{\sum}
            \gamma_{t_{j'}} \gamma_{t_j}
            \ket{t_j}_2 \ket{j'}_4 
        \right)\\
        &\qquad \qquad + \left(
            \sum_{n \notin \mathcal{C}_2} \gamma_n \ket{n}_2 \ket{k_2}_4
            +
            \sum_{j'|t_{j'} \in \mathcal{C}_2} \gamma_{t_{j'}} \ket{\text{vac}}_2 \ket{j',k_2}_4
        \right)
    \end{aligned}
    \right].
\end{equation}

As can be seen the probability for each mode to not contain the vacuum state after the operation of the SMR is $1/2$. Therefore, for two SMRs acting in parallel with each other, the probability for every mode to not contain the vacuum state is $1/4$. Moreover, by definition, the preparation of ancilla state requires us to construct the states $\ket{\xi_i}$ which in turn depend on the input qudit states. As such, a partial swap that does not change the state of the input qudits requires a priori knowledge about the input qudit states, making the overall scheme state-dependent. Moreover, we note that for a single trigger mode in each qudit, the ancilla can be prepared without the knowledge of the input states.

The joint state after post-selecting on the non-vacuum states is
\begin{equation}
\begin{aligned}
\ket{\psi}_{1234}
= & \frac{1}{\sqrt{P_1 P_2}}\underset{\substack{c_i \in \mathcal{C}_1 , t_j \in \mathcal{C}_2 \\ i'|c_{i'} \in \mathcal{C}_1 \\ j'|t_{j'} \in \mathcal{C}_2}}{\sum}
 \beta_{c_i}
 \gamma_{t_j}
 \beta_{c_{i'}}
 \gamma_{t_{j'}}
 \ket{c_i}_1 \ket{t_j}_2  \ket{i'}_3  \ket{j'}_4
 + \frac{1}{\sqrt{P_1}}\underset{\substack{c_i \in \mathcal{C}_1 \\ n \notin \mathcal{C}_2, i'|c_{i'} \in \mathcal{C}_1}}{\sum}
\beta_{c_{i}} \gamma_n \beta_{c_{i'}}
\ket{c_{i}}_1 \ket{n}_2 \ket{i'}_3 \ket{k_2}_4 \\   
&\quad + \frac{1}{\sqrt{P_2}} \underset{\substack{ m \notin \mathcal{C}_1 \\t_{j} \in \mathcal{C}_2, j' \mid c_{j'}\in \mathcal{C}_2}}{\sum}
\beta_{m} \gamma_{t_{j}} \gamma_{t_{j'}}
\ket{m}_1 \ket{t_j}_2 \ket{k_1}_3 \ket{j'}_4
+\underset{\substack{m \notin \mathcal{C}_1 \\ n \notin \mathcal{C}_2}}{\sum}
\beta_m \gamma_n
\ket{m}_1 \ket{n}_2 \ket{k_1}_3 \ket{k_2}_4.
\end{aligned}
\label{eq:SMR_state}
\end{equation}

The successful operation is heralded by a coincidence detection, where we post-select for events yielding exactly one photon click per system.

\section{Unitary transformations $O_1$ and $O_2$}
\label{appendix:unitaries}

Following the action of SMRs, we transform the ancilla qudits via unitary transformations $O_1$ and $O_2$ defined as
\begin{equation}
O_i = \ket{0}\bra{k_i}_{i+2} + \ket{1}\bra{\xi_{i}}_{i+2} + \sum_{j=2}^{k_i} \ket{j}\bra{v^{(i)}_j}_{i+2}, i \in \{1,2\},
\label{eq:merging}
\end{equation}
where the states $\lbrace\ket{v^{(i)}_j}\rbrace_{j\geq2}$ complete the orthonormal basis for the $k_i + 1$ dimensional system. These can be constructed using Gram-Schmidt orthonormalization with respect to the states $\ket{k_i}$ and $\ket{\xi_i}$. Physically, $O_i$ maps the states $\ket{k_i}$ and  $\ket{\xi_i}$ onto $\ket{0}$ and $\ket{1}$ respectively, while mapping the remaining orthogonal states $\{\ket{v_j}\}_{j\geq 2}$ onto $\{\ket{j}\}_{j \ge 2}$. 

Here, it should be noted that the transformations $O_i$ can be optically realized using Reck and Clements multiport interferometer constructions~\cite{RZB94, WPB16}. Furthermore, by definition, the transformations $O_i$ explicitly depend on the ancilla qudit states $\ket{\xi_i}_{i+2}$, which in turn depend on the input qudit states. Moreover, we note that for single trigger modes in each input qudit, such a requirement is not needed and for these special instances, the scheme is state independent.

The joint quantum state~\eqref{eq:SMR_state} after the action of the SMRs can be re-written as 
\begin{equation}
\begin{aligned}
\ket{\psi}_{1234}
= {}&
\underset{\substack{c_i \in \mathcal{C}_1 ,\, t_j \in \mathcal{C}_2}}{\sum}
 \beta_{c_{i}}
 \gamma_{t_j}
 \ket{c_i}_1 \ket{t_j}_2
 \left( \frac{1}{\sqrt{P_1}}\sum_{i' \mid c_{i'} \in \mathcal{C}_1}  \beta_{c_i}\ket{i'}_3 \right)
 \left(\frac{1}{\sqrt{P_2}} \sum_{j' \mid t_j \in \mathcal{C}_2} \beta_{t_j} \ket{j'}_4 \right) \\
&\quad
+ \underset{\substack{c_i \in \mathcal{C}_1,\, n \notin \mathcal{C}_2}}{\sum}
\beta_{c_{i}} \gamma_n 
\ket{c_{i}}_1 \ket{n}_2
\left(\frac{1}{\sqrt{P_1}} \sum_{i' \mid c_{i'} \in \mathcal{C}_1} \beta_{c_{i'}} \ket{i'}_3 \right)
\ket{k_2}_4 \\
&\quad
+ \underset{\substack{ m \notin \mathcal{C}_1,\, t_{j} \in \mathcal{C}_2}}{\sum}
\beta_{m} \gamma_{t_j}
\ket{m}_1 \ket{t_j}_2
\ket{k_1}_3
\left( \frac{1}{\sqrt{P_2}} \sum_{j' \mid t_{j'} \in \mathcal{C}_2}  \gamma_{t_{j'}} \ket{j'}_4 \right) \\
&\quad
+ \underset{\substack{m \notin \mathcal{C}_1 \\ n \notin \mathcal{C}_2}}{\sum}
\beta_m \gamma_n
\ket{m}_1 \ket{n}_2 \ket{k_1}_3 \ket{k_2}_4
\\[1ex]
= {}&
\underset{\substack{c_i \in \mathcal{C}_1 ,\, t_j \in \mathcal{C}_2}}{\sum}
 \beta_{c_{i}}
 \gamma_{t_j}
 \ket{c_i}_1 \ket{t_j}_2 \ket{\xi_1}_3 \ket{\xi_2}_4 + \underset{\substack{c_i \in \mathcal{C}_1,\, n \notin \mathcal{C}_2}}{\sum}
\beta_{c_{i}} \gamma_n 
\ket{c_{i}}_1 \ket{n}_2
\ket{\xi_1}_3 \ket{k_2}_4 \\
&\quad
+ \underset{\substack{ m \notin \mathcal{C}_1,\, t_{j} \in \mathcal{C}_2}}{\sum}
\beta_{m} \gamma_{t_j}
\ket{m}_1 \ket{t_j}_2
\ket{k_1}_3
\ket{\xi_2}_4 + \underset{\substack{m \notin \mathcal{C}_1 \\ n \notin \mathcal{C}_2}}{\sum}
\beta_m \gamma_n
\ket{m}_1 \ket{n}_2 \ket{k_1}_3 \ket{k_2}_4
\end{aligned}
\end{equation}
where in the second equality we have used Eq.~\eqref{eq:ancillas1}. 

The joint action of $O_1$ and $O_2$ on the ancilla qudits $3$ and $4$ is then given by
\begin{equation}
\begin{aligned}
\left( \mathds{1} \otimes \mathds{1} \otimes O_1 \otimes O_2 \right)
\ket{\psi}_{1234}
=&
\underset{\substack{c_i \in \mathcal{C}_1 \\ t_j \in \mathcal{C}_2}}{\sum}
\beta_{c_i}\gamma_{t_j}
\ket{c_i}_1 \ket{t_j}_2 \ket{1}_3 \ket{1}_4 +
\underset{\substack{c_i \in \mathcal{C}_1 \\ n \notin \mathcal{C}_2}}{\sum}
\beta_{c_i}\gamma_n
\ket{c_i}_1 \ket{n}_2 \ket{1}_3 \ket{0}_4
\\
&+
\underset{\substack{m \notin \mathcal{C}_1 \\ t_j \in \mathcal{C}_2}}{\sum}
\beta_m\gamma_{t_j}
\ket{m}_1 \ket{t_j}_2 \ket{0}_3 \ket{1}_4 +
\underset{\substack{m \notin \mathcal{C}_1 \\ n \notin \mathcal{C}_2}}{\sum}
\beta_m\gamma_n
\ket{m}_1 \ket{n}_2 \ket{0}_3 \ket{0}_4 .
\end{aligned}
\end{equation}

For notational consistency, we switch to the original indices using $c_i = m \text{ } \forall c_i \in \mathcal{C}_1$ and $t_j = n \text{ } \forall t_j \in \mathcal{C}_2$. We can write
\begin{equation}
\begin{aligned}
\left(\mathds{1} \otimes \mathds{1} \otimes O_1 \otimes O_2 \right)\ket{\psi}_{1234} =
&\underset{\substack{m \in \mathcal{C}_1 \\ n \in \mathcal{C}_2}}{\sum}
 \beta_{m}
 \gamma_{n}
 \ket{m}_1 \ket{n}_2 \ket{1}_3 \ket{1}_4 + \underset{\substack{m \in \mathcal{C}_1 \\ n \notin \mathcal{C}_2}}{\sum}
\beta_{m} \gamma_n
\ket{m}_1 \ket{n}_2 \ket{1}_3 \ket{0}_4
\\
&\qquad  +
\underset{\substack{m \notin \mathcal{C}_1 \\ n \in \mathcal{C}_2}}{\sum}
\beta_m \gamma_{n}
\ket{m}_1 \ket{n}_2 \ket{0}_3 \ket{1}_4 +
\underset{\substack{m \notin \mathcal{C}_1 \\ n \notin \mathcal{C}_2}}{\sum}\beta_m \gamma_n
\ket{m}_1 \ket{n}_2 \ket{0}_3 \ket{0}_4\\
&= \left(\sum_{m\notin \mathcal{C}_1}\beta_m \ket{m}_1\ket{0}_3 + \sum_{m \in \mathcal{C}_1} \beta_m \ket{m}_1\ket{1}_3\right) \left(\sum_{n \notin \mathcal{C}_1} \gamma_n \ket{n}_2\ket{0}_4 + \sum_{n \in \mathcal{C}_2} \gamma_n \ket{n}_2\ket{1}_4\right),
\end{aligned}
\label{eq:Os}
\end{equation}
where in the last equality, we note that the systems $1$, $3$ and $2$, $4$ are in an entangled state.

\section{Bell state measurement}
\label{appendix:BSM}

Following the operations of $O_1$ and $O_2$, a Hadamard gate is applied only to the qudit $4$ to obtain
\begin{equation}
\ket{\Psi}_ = \frac{1}{\sqrt{2}} 
\left[
\begin{aligned}
&\underset{\substack{m \in \mathcal{C}_1 \\ n \in \mathcal{C}_2}}{\sum}
 \beta_{m}
 \gamma_{n}
 \ket{m}_1 \ket{n}_2 \left( \ket{1}_3 \ket{0}_4 - \ket{1}_3 \ket{1}_4 \right) + \underset{\substack{m \in \mathcal{C}_1 \\ n \notin \mathcal{C}_2}}{\sum}
\beta_{m} \gamma_n
\ket{m}_1 \ket{n}_2 \left( \ket{1}_3 \ket{0}_4 + \ket{1}_3 \ket{1}_4 \right)
\\
&\qquad + 
\underset{\substack{m \notin \mathcal{C}_1 \\ n \in \mathcal{C}_2}}{\sum}
\beta_m \gamma_{n}
\ket{m}_1 \ket{n}_2 \left( \ket{0}_3 \ket{0}_4 - \ket{0}_3 \ket{1}_4 \right) +
\underset{\substack{m \notin \mathcal{C}_1 \\ n \notin \mathcal{C}_2}}{\sum}
\beta_m \gamma_n
\ket{m}_1 \ket{n}_2 \left( \ket{0}_3 \ket{0}_4 + \ket{0}_3 \ket{1}_4 \right)
\end{aligned}
\right],
\label{eq:Hadamard}
\end{equation}
which can be re-written as
\begin{equation}
\ket{\Psi} = \frac{1}{2} 
\left[
\begin{aligned}
&\underset{\substack{m \notin \mathcal{C}_1 \\ n \notin \mathcal{C}_2}}{\sum}
    \beta_{m} \gamma_n  \ket{m}_1 \ket{n}_2 \left(\ket{\phi^+}_{34} + \ket{\phi^-}_{34} + \ket{\psi^+}_{34} + \ket{\psi^-}_{34}\right) \\
    &\qquad +
    \underset{\substack{m \notin \mathcal{C}_1 \\ n \in \mathcal{C}_2}}{\sum}
    \beta_{m} \gamma_n\ket{m}_1 \ket{n}_2 \left(\ket{\phi^+}_{34} + \ket{\phi^-}_{34} - \ket{\psi^+}_{34} - \ket{\psi^-}_{34}\right) \\
    &\qquad\qquad  + \underset{\substack{m \in \mathcal{C}_1 \\ n \notin \mathcal{C}_2}}{\sum}
    \beta_{m} \gamma_n \ket{m}_1 \ket{n}_2 \left(\ket{\phi^+}_{34} -   \ket{\phi^-}_{34} + \ket{\psi^+}_{34} - \ket{\psi^-}_{34}\right) \\
    &\qquad\qquad\qquad + \underset{\substack{m \in \mathcal{C}_1 \\ n \in \mathcal{C}_2}}{\sum} \beta_{m} \gamma_n \ket{m}_1 \ket{n}_2 \left(-\ket{\phi^+}_{34} +\ket{\phi^-}_{34} + \ket{\psi^+}_{34} - \ket{\psi^-}_{34}\right)
\end{aligned}
\right],
\end{equation}
where $\lbrace \ket{\phi^+}_{34}, \ket{\phi^-}_{34}, \ket{\psi^+}_{34}, \ket{\psi^-}_{34} \rbrace$ are the four Bell states
\begin{equation}
\begin{aligned}
    \ket{\phi^{\pm}}_{34} = \frac{1}{\sqrt{2}}\left( \ket{0}_3\ket{0}_4 \pm \ket{1}_3\ket{1}_4 \right), \\
    \ket{\psi^{\pm}}_{34} = \frac{1}{\sqrt{2}}\left( \ket{0}_3\ket{1}_4 \pm \ket{1}_3\ket{0}_4 \right).
\end{aligned}
\end{equation}

By grouping the terms together, the joint state of the four qudits can be rewritten as:
\begin{equation}
\begin{aligned}
\begin{aligned}
\begin{aligned}
\ket{\Psi}
&=\frac{1}{2}\left(
\begin{aligned}
    \underset{\substack{m \notin \mathcal{C}_1 \\ n \notin \mathcal{C}_2}}{\sum} \beta_{m} \gamma_n \ket{m}_1 \ket{n}_2
    + \underset{\substack{m \notin \mathcal{C}_1 \\ n \in \mathcal{C}_2}}{\sum}
    \beta_{m} \gamma_n \ket{m}_1\ket{n}_2
\\
    + \underset{\substack{m \in \mathcal{C}_1 \\ n \notin \mathcal{C}_2}}{\sum}
    \beta_{m} \gamma_n\ket{m}_1 \ket{n}_2 -
    \underset{\substack{m \in \mathcal{C}_1 \\ n \in \mathcal{C}_2}}{\sum}
    \beta_{m} \gamma_n  \ket{m}_1 \ket{n}_2
\end{aligned}
\right) \ket{\phi^+}_{34}
+
\frac{1}{2}\left(
\begin{aligned}
    \underset{\substack{m \notin \mathcal{C}_1 \\ n \notin \mathcal{C}_2}}{\sum} \beta_{m} \gamma_n \ket{m}_1 \ket{n}_2
    + \underset{\substack{m \notin \mathcal{C}_1 \\ n \in \mathcal{C}_2}}{\sum}
    \beta_{m} \gamma_n \ket{m}_1\ket{n}_2
\\
    - \underset{\substack{m \in \mathcal{C}_1 \\ n \notin \mathcal{C}_2}}{\sum}
    \beta_{m} \gamma_n\ket{m}_1 \ket{n}_2 +
    \underset{\substack{m \in \mathcal{C}_1 \\ n \in \mathcal{C}_2}}{\sum}
    \beta_{m} \gamma_n  \ket{m}_1 \ket{n}_2
\end{aligned}
\right) \ket{\phi^-}_{34}
\\
&+
\frac{1}{2}\left(
\begin{aligned}
    \underset{\substack{m \notin \mathcal{C}_1 \\ n \notin \mathcal{C}_2}}{\sum} \beta_{m} \gamma_n \ket{m}_1 \ket{n}_2
    - \underset{\substack{m \notin \mathcal{C}_1 \\ n \in \mathcal{C}_2}}{\sum}
    \beta_{m} \gamma_n \ket{m}_1\ket{n}_2
\\
    + \underset{\substack{m \in \mathcal{C}_1 \\ n \notin \mathcal{C}_2}}{\sum}
    \beta_{m} \gamma_n\ket{m}_1 \ket{n}_2 +
    \underset{\substack{m \in \mathcal{C}_1 \\ n \in \mathcal{C}_2}}{\sum}
    \beta_{m} \gamma_n  \ket{m}_1 \ket{n}_2
\end{aligned}
\right) \ket{\psi^+}_{34}
+
\frac{1}{2}\left(
\begin{aligned}
    \underset{\substack{m \notin \mathcal{C}_1 \\ n \notin \mathcal{C}_2}}{\sum} \beta_{m} \gamma_n \ket{m}_1 \ket{n}_2
    - \underset{\substack{m \notin \mathcal{C}_1 \\ n \in \mathcal{C}_2}}{\sum}
    \beta_{m} \gamma_n \ket{m}_1\ket{n}_2
\\
    - \underset{\substack{m \in \mathcal{C}_1 \\ n \notin \mathcal{C}_2}}{\sum}
    \beta_{m} \gamma_n\ket{m}_1 \ket{n}_2 -
    \underset{\substack{m \in \mathcal{C}_1 \\ n \in \mathcal{C}_2}}{\sum}
    \beta_{m} \gamma_n  \ket{m}_1 \ket{n}_2
\end{aligned}
\right) \ket{\psi^-}_{34}
\end{aligned}
\end{aligned}
\end{aligned}.
\end{equation}

The aforementioned state can be further simplified as
\begin{equation}
    \ket{\Psi} = \frac{1}{2} \left(
    U_{\text{MCZ}}\ket{\psi}_{12}\ket{\phi^+}_{34} + U_1U_{\text{MCZ}}\ket{\psi}_{12}\ket{\phi^-}_{34} + U_2U_{\text{MCZ}}\ket{\psi}_{12}\ket{\psi^+}_{34} + U_1U_2U_{\text{MCZ}}\ket{\psi}_{12}\ket{\psi^-}_{34}
    \right),
\end{equation}
where $\ket{\psi}_{12} = \ket{\psi'}_1 \ket{\psi''}_2$
and
\begin{equation}
    U_{1} = \mathds{1} - 2\ket{\xi_{1}}\bra{\xi_{1}}_1, \quad
    U_{2} = \mathds{1} - 2\ket{\xi_{2}}\bra{\xi_{2}}_2.
\end{equation}

The outcome of a BSM projects the two input qudits to the desired state up to local unitaries $U_1$ and $U_2$. Based on the observed outcome of the BSM, an appropriate unitary transformation can then be applied to realize the operation $U_{\text{MCZ}}$.

\section{State-independent approach}
\label{appendix:state_independent}

The scheme to implement the operation $U_{\text{MCZ}} = \mathds{1}-2\underset{\substack{m\in\mathcal{C}_1\\ n\in \mathcal{C}_2}}{\sum}\ket{m}\bra{m}_1 \otimes \ket{n}\bra{n}_2$ with $m$, $n$ number of trigger modes in the first and second qudit relies on implementing the SMRs on input qudit and ancilla qudit pairs which effectively flag the state of the input qudits as a trigger (or a non-trigger) state by swapping the trigger modes $\{c_i,t_j\}$ with the ancilla modes $\{i,j\}$ followed by post-selection (see Eq.~\eqref{eq:smr_operation_appendix}). Specifically, if the input qudit occupies a trigger mode $\{c_i\}$ ($\{t_j\}$) then the ancilla qudit occupies the corresponding mode $i$ ($j$). Similarly, if the input qudit occupies a non-trigger mode then the ancilla will occupy the additional mode $k_i$ ($i \in \{1,2\}$). As a result, the joint states of input qudit ancilla constitutes an entangled state. Since the ancilla effectively carries a binary information (whether or not the input qudit occupies the trigger mode), we benefit from mapping the ancillae to two-dimensional systems (qubits) through local unitary operations $O_1$ and $O_2$, ultimately constructing the desired resource state (Eq.~\eqref{eq:Os}). However, the partial swap operations of the SMRs change the state of input qudits unless the swapped information between the input qudit and the ancilla is exactly the same. Since we cannot make copies of the input qudits to facilitate this process because of no-cloning theorem, we rely on a priori knowledge of the input qudit states to prepare the desired ancilla state. Moreover, it is due to this knowledge that we can design $O_1$ and $O_2$. However, for $|\mathcal{C}_i| = 1$, i.e., single trigger modes for each input qudit, the ancilla state preparation is a trivial process because $\ket{\xi_0} = \ket{c_0}$ and $\ket{\xi_1} = \ket{t_0}$ for $\mathcal{C}_1 = \{c_0\} \text{ and } \mathcal{C}_2 = \{t_0\}$ as the sets of trigger modes for the two qudits (see Eq.~\eqref{eq:ancillas1}). Moreover, $k_i+1=2$ (where $k_i = |\mathcal{C}_i|$ for $i \in \{1,2\}$), which implies that the ancillae are two-dimensional systems due to which $O_1$ and $O_2$ act as identity. As such for this particular instance, no a priori knowledge about the input qudit states is necessary, making the overall scheme state independent.

Using this fact, it is possible to design a state independent realization of the $U_{\text{MCZ}}$ operation with multiple number of trigger modes in the first and second qudit. Qualitatively, this realization proceeds by introducing qubit ancillas, rather than qudit ancillas, and sequentially applying $k_1$ instances of the proposed multi-level CZ gate, each conditioned on a single trigger mode of the input qudit. An analogous sequence of $k_2$ such gates is applied to the second qudit.

Consider the two-level CZ gates on input qudit $1$ and ancilla qubit $3$ defined by the action
\begin{equation} 
U_{\text{MCZ}}^s=\mathds{1}-2\ket{s}\bra{s}_1\otimes \ket{1}\bra{1}_3,
\end{equation}
where $s \in \mathcal{C}_1$ corresponds to a trigger mode. We can then define the operation $\Tilde{O}_1$ on the input qudit state $\ket{\psi'}_1$ and ancilla qubit initialized in the state $\ket{0}_3$ (without loss of generality) as
\begin{equation}
\begin{aligned}
\Tilde{O}_1 \ket{\psi'}_1\ket{0}_3 =
&\left(\mathds{1} \otimes H\right)\left(\prod_{s = 0}^{k_1-1} U_{\text{MCZ}}^{s}\right) \left(\mathds{1} \otimes H\right) \left(\sum_{m\notin \mathcal{C}_1}\beta_m\ket{m}_1\ket{0}_3+\sum_{m\in \mathcal{C}_1}\beta_m\ket{m}_1\ket{0}_3\right) \\
=& \left(\mathds{1} \otimes H\right)\left(\prod_{s = 0}^{k_1-1} U_{\text{MCZ}}^{s}\right) \frac{1}{\sqrt{2}}\left[\sum_{m\notin  \mathcal{C}_1}\beta_m\ket{m}_1\left(\ket{0}_3+\ket{1}_3\right)+\sum_{m\in \mathcal{C}_1}\beta_m\ket{m}_1\left(\ket{0}_3+\ket{1}_3\right)\right].
\end{aligned}
\end{equation}
where in first equality we have used Eq.~\eqref{eq:definitions} and decomposed the summation over the indices $m\in \mathcal{C}_1$ and $m \notin \mathcal{C}_1$. After sequentially applying $U_{\text{MCZ}}^{s}$ $k_1$ times with the indices $s \in \mathcal{C}_1$ we obtain
\begin{equation}
\begin{aligned}
\Tilde{O}_1 \ket{\psi'}_1\ket{0}_3 = &\left(\mathds{1} \otimes H\right) \frac{1}{\sqrt{2}}\left[\sum_{m\notin \mathcal{C}_1}\beta_m\ket{m}_1\left(\ket{0}_3+\ket{1}_3\right) + \sum_{m\in \mathcal{C}_1}\beta_m\ket{m}_1\left(\ket{0}_3-\ket{1}_3\right)\right] \\
=&\sum_{m\notin \mathcal{C}_1}\beta_m\ket{m}_1\ket{0}_3 + \sum_{m\in \mathcal{C}_1}\beta_m\ket{m}_1\ket{1}_3,
\end{aligned}
\end{equation}
which gives the desired entangled state between the input qudit and the ancilla.

Similarly for $\ket{\psi''}_2\ket{0}_4$ we get,
\begin{equation}
\Tilde{O}_2 \ket{\psi''}_2\ket{0}_4 =  \sum_{n\notin \mathcal{C}_2}\gamma_n\ket{n}_2\ket{0}_4 + \sum_{n\in \mathcal{C}_2}\gamma_n\ket{n}_2\ket{1}_4.
\end{equation}

The joint state of the four systems then gives the resource state as in Eq.~\eqref{eq:Os}. After the application of $\tilde{O}_1$ and $\tilde{O}_2$ we then proceed with the BSM similar to the state-dependent scheme. The probability of successful optical realization of this scheme scales as $\frac{1}{2}(\frac{1}{8})^{k_1+k_2}$, where the factor $1/2$ arises because of the probabilistic nature of BSM in LOQC.

\twocolumngrid

%

\end{document}